\documentclass[a4paper,11pt]{article}
\pdfoutput=1
\usepackage{jheppub}
\usepackage{setspace}
\usepackage[caption=false]{subfig}
\usepackage{graphicx}
\usepackage{dcolumn}
\usepackage{bm}
\usepackage{mathtools}
\usepackage{amsthm}

\usepackage{color}
\usepackage{hyperref}
\hypersetup{pdftitle={Rescuing Complementarity With Little Drama}, pdfauthor={Ning Bao, Adam Bouland, Aidan Chatwin-Davies, Jason Pollack, Henry Yuen}, citecolor=blue,linkcolor=blue,urlcolor=blue,citecolor=blue}

\newcommand{\hil}{\mathcal{H}}

\newtheorem{theorem}{Theorem}

\theoremstyle{definition}

\newtheorem{conjecture}[theorem]{Conjecture}

\newcommand{\ket}[1]{{|#1\rangle}}

\newcommand{\ketbra}[2]{|#1\rangle\! \langle #2|}
\newcommand{\Tr}{\mbox{\rm Tr}}

\newcommand{\eps}{\varepsilon}

\newcommand{\Eq}[1]{Eq.~(\ref{#1})}
\newcommand{\Eqs}[2]{Eqs.~(\ref{#1}) and (\ref{#2})}
\newcommand{\Sec}[1]{Sec.~\ref{#1}}

\newcommand{\Fig}[1]{Fig.~\ref{#1}}

\preprint{CALT-TH-2016-017}

\title{Rescuing Complementarity With Little Drama}

\author[a]{Ning Bao}
\author[b]{Adam Bouland}
\author[a]{Aidan Chatwin-Davies}
\author[a]{Jason Pollack}
\author[c]{Henry Yuen}

\affiliation[a]{
Walter Burke Institute for Theoretical Physics, California Institute of Technology,\\
1200 East California Boulevard,
Pasadena, CA 91125, U.S.A.
}
\affiliation[b]{
Computer Science and Artificial Intelligence Laboratory, Massachusetts Institute of Technology,\\
77 Massachusetts Avenue, Cambridge, MA 02139, U.S.A.}
\affiliation[c]{
Computer Science Division, University of California, Berkeley,\\
Berkeley, CA 94720, U.S.A.
}

\emailAdd{ningbao@caltech.edu}
\emailAdd{adam@csail.mit.edu}
\emailAdd{achatwin@caltech.edu}
\emailAdd{jpollack@caltech.edu}
\emailAdd{hyuen@cs.berkeley.edu}

\abstract{The AMPS paradox challenges black hole complementarity by apparently constructing a way for an observer to bring information from the outside of the black hole into its interior if there is no drama at its horizon, making manifest a violation of monogamy of entanglement. 
We propose a new resolution to the paradox: this violation cannot be explicitly checked by an infalling observer in the finite proper time they have to live after crossing the horizon.
Our resolution depends on a weak relaxation of the no-drama condition (we call it ``little-drama'') which is the ``complementarity dual'' of scrambling of information on the stretched horizon. 
When translated to the description of the black hole interior, this implies that the fine-grained quantum information of infalling matter is rapidly diffused across the entire interior while classical observables and coarse-grained geometry remain unaffected.
Under the assumption that information has diffused throughout the interior, we consider the difficulty of the information-theoretic task that an observer must perform after crossing the event horizon of a Schwarzschild black hole in order to verify a violation of monogamy of entanglement.
We find that the time required to complete a necessary subroutine of this task, namely the decoding of Bell pairs from the interior and the late radiation, takes longer than the maximum amount of time that an observer can spend inside the black hole before hitting the singularity.
Therefore, an infalling observer cannot observe monogamy violation before encountering the singularity.}

\begin{document}
\maketitle
\flushbottom

\section{Introduction}

The information paradox \cite{Hawking:2005kf} and its more modern AMPS incarnation \cite{Almheiri:2012rt, Almheiri:2013hfa} are deeply puzzling issues lying at the center of any attempts at reconciling quantum mechanics with gravity. 
Black hole complementarity, as proposed by \cite{Susskind:1993if}, attempted to resolve the information paradox by asserting that information that falls into the black hole interior is also retained at the stretched horizon.
Observers are only able to access this information in one of two ``complementary'' descriptions, either in the interior or at the horizon, so that the apparent violation of the no-cloning theorem visible in a global description could never be verified.
AMPS, however, considered a scenario in which an observer first collects information on the outside by gathering Hawking radiation, then jumps through the horizon and into the black hole interior.
Assuming standard postulates of black hole complementarity, namely
\begin{enumerate}
\item unitarity,
\item the validity of low-energy effective field theory outside the stretched horizon,
\item that the black hole is a quantum mechanical system with dimension given by $e^{A/4}$,
\end{enumerate}
and further
\begin{enumerate}
\item[4.] that the horizon is not a special place---that ``no drama" happens at the horizon, so an observer can actually enter the black hole interior,
\end{enumerate}
AMPS pointed out an apparent violation of monogamy of entanglement\footnote{Monogamy of entanglement is the statement that no single qubit can be simultaneously maximally entangled with two different systems.} among three systems: the black hole interior, the recently emitted Hawking radiation (late radiation), and the previously emitted Hawking radiation (early radiation).
To avoid this violation, it therefore seemed necessary to give up one of the assumptions mentioned above, all of which are cherished pillars of modern physics.
Giving up the final assumption would mean that observers who attempt to enter the black hole would be violently destroyed by high-energy excitations, hence the name ``firewall paradox."

This led to a flurry of attempts to resolve the paradox by weakening one or more of the core axioms, or by changing the paradigm completely \cite{Chen:2014jwq, Maldacena:2013xja, Lloyd:2013bza, Papadodimas:2012aq, Mathur:2005zp, Giddings:2012gc,Hotta:2015,Nomura:2014woa,Nomura:2014voa,Nomura:2016qum}.
Reaching consensus as to which resolution is the correct one has proven challenging.

An interesting proposed resolution to the information paradox, based on arguments from computational complexity, was given by Harlow and Hayden \cite{Harlow:2013tf}. They argued that the part of the AMPS experiment where the experimenter has to decode\footnote{To ``decode the entanglement" of a state $\ket{\psi}_{AB}$ is to act with local unitaries on $A$ and $B$ to create a Bell pair across $A$ and $B$. This is similar to the notion of entanglement distillation \cite{bennettdistillation1996}, except here we have only one copy of the state $\ket{\psi}_{AB}$, whereas in distillation one has multiple identical copies of the state.
} entanglement between the old radiation and the late radiation of the black hole involves an extremely difficult computational task. Under very plausible conjectures in computational complexity\footnote{Namely, that quantum computers cannot efficiently invert \emph{cryptographic one-way functions}.}, the time required to perform this quantum computation in general would be exponentially longer than the evaporation time of the black hole. Thus, by the time that the entanglement is decoded, there will remain no black hole within which to check for the violation of monogamy of entanglement. While the two quantum mechanical descriptions of the black hole appear to imply a violation of monogamy, this apparent violation cannot be ``revealed'' by the AMPS experiment, and thus the experimenter does not see any contradiction with quantum mechanics. Just like the original violation of no-cloning in black hole complementarity itself, this would signal that only the various partial descriptions accessible by a single observer should be considered. 

The main appeal of this argument is that it does not require a weakening of any of the core assumptions mentioned previously. However, it is not without its vulnerabilities. For example, Oppenheim and Unruh~\cite{Oppenheim:2014gfa} gave an argument showing that a very motivated experimenter could evade the Harlow-Hayden complexity barrier by offloading the hard computation into a ``precomputation'' phase before the black hole had even formed, and then perform the AMPS experiment efficiently using the ``cached computation.'' Another vulnerability is that the computational hardness of the Harlow-Hayden argument assumes that the black hole in question somehow encodes a cryptographically difficult one-way function; however, one may be able to set up a black hole so that the entanglement decoding task is particularly easy \cite{AaronsonPC}.

Nevertheless, the Harlow-Hayden proposal remains a compelling one, and it sets the context for the argument that we present in this paper. Here, we also study whether ideas from information theory and computer science can help resolve the information paradox, but in another setting: whereas Harlow and Hayden focus on the computational complexity of the AMPS experiment \emph{outside} the black hole, we examine the information processing that must be performed \emph{inside} the black hole in order to check for violations of monogamy of entanglement. This is a potentially different line of argument, because while it might be possible to evade computational limits outside of the horizon \cite{Almheiri:2013hfa,Oppenheim:2014gfa}, one certainly cannot extend one's time inside the horizon, as an infalling observer invariably hits the singularity in a bounded amount of time.

In this paper we study an observer who begins outside of an evaporating Schwarzschild black hole well after the Page time and who has learned that a subset of late Hawking radiation that she holds is maximally entangled with the early Hawking radiation\footnote{Though this is the task that Harlow and Hayden argue is difficult, we assume for the purpose of the argument that this task has been achieved.}.
We suppose that the observer then enters the black hole, sees no firewall, and then attempts to decode maximal entanglement between the late radiation that she holds and the black hole interior.
If she succeeds in completing this task, she can then perform measurements on an ensemble of her decoded Bell pairs in order to probabilistically detect a violation of monogamy of entanglement.
We compare the proper time it takes for the observer to perform this procedure with the infall time before the observer hits the singularity.
We find that, under the assumption that the subsystem of the black hole interior with which the observer's late radiation is entangled has diffused throughout the whole interior at the time she crosses the horizon, the observer will not have enough time to complete even the first step of the procedure, i.e., entanglement decoding, before encountering the singularity. As such, while a global description, if it existed, would contain an implicit violation of monogamy of entanglement, an observer who entered the black hole would unable to directly verify any such violation.
Therefore, our resolution of the firewalls paradox is similar in spirit to complementarity \cite{Susskind:1993if} in the sense that apparent global violations of quantum mechanics are not verifiable by local observers.

The assumption that we make about dynamics inside the horizon is a mild weakening of the no-drama condition that is typically considered: while we expect no-drama to hold for macroscopic, classical objects that cross the
event horizon, fine-grained quantum information should be \emph{scrambled} throughout the black hole's degrees of freedom, regardless of whether these degrees of freedom are described as the black hole horizon or as the black hole interior. In particular, the assertion that an observer inside the black hole sees such scrambling is the novel assumption of our paper. We thus call this assumption ``little drama,'' and it is central to our argument.

The organization of this paper is as follows.
In Section II, we review facts about black holes and their scrambling from the perspective of different observers in spacetime.
In Section III, we focus on the specific task of collecting a late-time Hawking radiation particle, assess the degree of scrambling that has occurred prior to the observer crossing the stretched horizon of the black hole, and give a discussion of the little-drama condition.
In Section IV, we combine all the ingredients from the previous sections and analyze the time needed to perform the task of checking for violations of monogamy. Finally, we discuss and conclude in Sections V and VI.

\section{Background: Black Holes and Scrambling} \label{sec:scrambling}

In the thought experiments to follow, we will consider black holes that formed from the gravitational collapse of matter and that eventually evaporate into a gas of Hawking radiation. 
We will assume that the initial mass of any black hole that we consider is large enough that physics outside the black hole is well-described by effective field theory on a black hole background in regions of spacetime that are sufficiently distant from the end of evaporation.
We will also suppose that the process of black hole formation and evaporation is a fundamentally unitary process.
As such, if the matter that collapsed to form a black hole was initially in a pure quantum state, then the state of the Hawking radiation after evaporation---as well as any combined intermediate state of the black hole and hitherto emitted Hawking radiation---is also a pure state.

\begin{figure}
\centering
\includegraphics[scale=0.75]{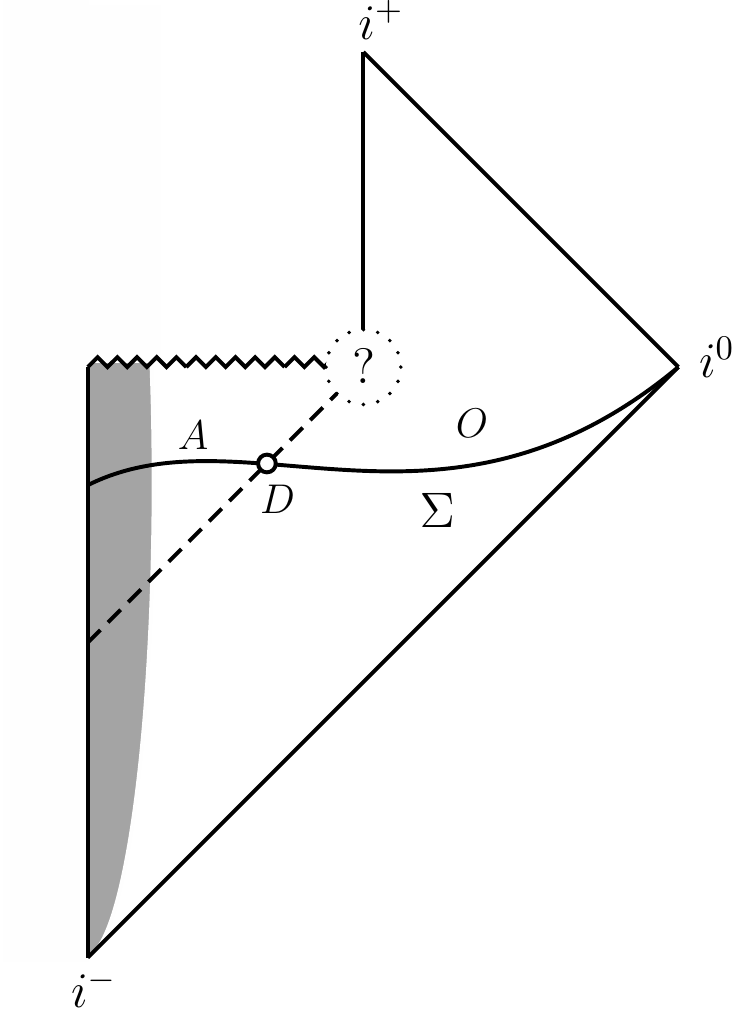}
\caption{Penrose diagram of a black hole that forms from the gravitational collapse of matter and that ultimately evaporates.}
\label{fig:bhpenrose}
\end{figure}

Consider now some observer who resides outside the black hole.
We will adopt the viewpoint that such an observer's observations are described according to complementarity \cite{Susskind:1993if} and the membrane paradigm \cite{Thorne:1986iy}.
Explicitly, suppose that the black hole spacetime is foliated by some set of achronal (spacelike or null) surfaces with respect to which the observer performs field-theoretic calculations.
In accordance with complementarity, an observer outside the black hole should not associate a Hilbert space to an entire surface $\Sigma$ if it intersects the event horizon.
In such a case, she instead organizes the physical Hilbert space associated to $\Sigma$ into a tensor product $\hil = O \otimes D$.
The space $O$ describes the degrees of freedom on the portion of $\Sigma$ that lies outside of the black hole, and $D$ is a Hilbert space that describes the black hole's degrees of freedom and that is localized about the event horizon (\Fig{fig:bhpenrose}).
From the outside observer's point of view, all of physics is described by, and all processes play out in, these two Hilbert spaces; she never has to (and in fact \emph{may not}) make reference to the the black hole interior.\footnote{See also \cite{Nomura:2011dt} (in particular Sec.~4) as well as \Sec{sec:priorworks} for further discussion of the way in which $\mathcal{H}$ factorizes and the ways in which different factorizations are related as a consequence of assuming complementarity.}

We will suppose that $D$ is localized to the stretched horizon of the black hole \cite{Susskind:1993if}.
We take the outer boundary of the stretched horizon to be at a proper distance on the order of a Planck length above the event horizon.
As such, the outer boundary of the stretched horizon is a timelike surface with which an outside observer can interact.

Despite the fact that a complete theory of quantum gravity is not known and that the full dynamics of black holes are not understood, it is widely expected that the quantum state of matter gets scrambled when it enters the stretched horizon \cite{Hayden:2007cs, Sekino:2008he, Lashkari:2011yi}. There are many possible ways to define scrambling, but informally speaking, a system scrambles if it diffuses quantum information over all its degrees of freedom. In particular, a black hole has scrambled the information in a small subset $D^\prime \subset D$ when any initial entanglement between $D'$ and the outside $O$ gets distributed evenly throughout $D$, i.e., when almost all small subsets of $D$ have nearly the same amount of entanglement with $O$. After scrambling, an observer cannot recover this entanglement unless she examines a sizable fraction of the entire horizon $D$. 
 
The characteristic timescale over which scrambling occurs, called the scrambling time, is given by
\begin{equation}
t_s = \frac{1}{2\pi T} \ln \, S \, ,
\end{equation}
where $T$ and $S$ are the temperature and entropy of the black hole respectively \cite{Hayden:2007cs, Sekino:2008he, Shenker:2013pqa, Shenker:2014cwa, Maldacena:2015waa}.
(Both in this expression and throughout the paper we have set $c=k_\mathrm{B}=\hbar=1$.)
This time is measured with respect to the clock of an asymptotic observer who is far away from the black hole.
For example, for a Schwarzschild black hole in $3+1$ dimensions, the metric is given by
\begin{equation}
ds^2 = -\left(1-\frac{r_s}{r}\right) dt^2 + \left(1-\frac{r_s}{r}\right)^{-1} dr^2 + r^2 \, d\Omega_2^2,
\end{equation} the temperature is 
\begin{equation}
T=\frac{1}{8 \pi G M}=\frac{1}{4 \pi r_s},
\end{equation} and the entropy is
\begin{equation}
S=\frac{A}{4G}=\frac{4 \pi {r_s}^2}{4 l_P^2}=\frac{\pi {r_s}^2}{l_P^2}.
\end{equation} 
As such, the scrambling time is given by
\begin{equation} \label{eq:asympts}
t_s = {r_s} \ln \frac{\sqrt{\pi}{r_s}}{l_P} \, .
\end{equation}
The event horizon is located at $r={r_s}=2 G M$, and $l_P$ denotes the Planck length.
Importantly, a stationary observer who hovers at some fixed value of $r=r_0$ above the black hole sees scrambling happen faster, since her clock ticks faster relative to Schwarzschild time.
In other words, the scrambling time as measured in the proper time of a stationary observer at coordinate height $r_0$ is
\begin{equation}
\tau_s(r_0) = \sqrt{1-\frac{{r_s}}{r_0}} \, t_s \, .
\end{equation}
In particular, we can work out what the scrambling time at the stretched horizon must be.
If we fix the boundary of the stretched horizon to lie at a proper distance $l_P$ above the event horizon, one finds that this corresponds to a coordinate distance $r = {r_s} + \delta r$, where
\begin{equation} \label{eq:shorcoord}
\delta r = \frac{l_P^2}{4{r_s}} + O\left(\frac{l_P^3}{{r_s}^2}\right) \, .
\end{equation}
It then follows that
\begin{align}
\nonumber \tau_s({r_s}+\delta r) &= \sqrt{\frac{l_P^2}{l_P^2 + 4 {r_s}^2}} {r_s} \ln\left[\frac{\sqrt{\pi}{r_s}}{l_P}\right] \\[2mm]
&\approx \frac{l_P}{2}  \ln\left[\frac{\sqrt{\pi}{r_s}}{l_P}\right] \, , \label{eq:horizonts}
\end{align}
which is consistent with other calculations of the scrambling time at the stretched horizon \cite{Hayden:2007cs, Sekino:2008he}.

\section{Hawking radiation and scrambling: what Alice sees}

Having established the preliminaries, we can now begin to investigate the central question of this work: whether an observer who crosses the event horizon of an evaporating black hole can, in the absence of a firewall, verify a violation of monogamy of entanglement before she hits the singularity.
The answer to this question depends on several considerations: in particular, the nature of scrambling from the point of view of an observer inside the black hole, under what circumstance an ingoing Hawking mode is scrambled before an observer carrying the corresponding outgoing mode crosses the horizon, and the difficulty of undoing scrambling inside the black hole.
We address the first two points, the nature of scrambling and under what conditions scrambling occurs, in this section. In particular, we motivate the little-drama assumption used in the argument of this paper.

\subsection{Scrambling, inside and out}
\label{sec:innout}

Suppose that Alice has been monitoring a black hole since its formation and that she collects any Hawking radiation that it emits.
At some point well past the Page time, she decides to perform her ultimate experiment: an experimental test of the AMPS paradox.
To this end, she collects $k$ particles of (late) Hawking radiation and first checks whether they are maximally entangled with the radiation that was emitted earlier.
Let us momentarily grant Alice unlimited computational power outside of the black hole and suppose that she finds that these late quanta of radiation are indeed maximally entangled with the early radiation.
She then holds on to these final Hawking particles and enters the black hole.
To her transient relief, suppose that she does not encounter a firewall at the horizon.
As such, suspecting a possible violation of monogamy of entanglement, her next objective is to check whether the $k$ Hawking particles that she collected outside of the black hole are entangled with degrees of freedom in the black hole interior.

Recall that Hawking radiation consists of paired entangled excitations of field modes.
The outgoing modes constitute the radiation that is visible to stationary observers, but for each outgoing mode there is also an ingoing mode which remains confined to the black hole interior.
In principle, Alice's task is to ``catch up" with the ingoing excitations that correspond to the $k$ particles that she collected and check whether they are entangled.
In the next section, we will consider whether and how Alice can actually perform this check.
For now, we will consider a prerequisite question: what do the ingoing excitations look like to Alice should she catch up to them inside the black hole?

Because of complementarity, while Alice is outside of the black hole, she should not think of an ingoing excitation as some particle which falls toward the singularity.
Rather, she sees it as some excitation of the stretched horizon, which begins to scramble as the dynamics of the stretched horizon unfold.
Yet, also because of complementarity, Alice's description of physical processes changes once she crosses the event horizon of the black hole.
The stretched horizon is no more and she is now fully entitled to describe physics in the black hole interior.
For example, she can now associate a Hilbert space with each of her past lightcones and make the division $\hil =  A \otimes O$, where $A$ and $O$ describe degrees of freedom on the intersection of her past lightcone with the interior and exterior of the black hole respectively.
It is in this frame that she must look for the ingoing excitations.

Our aim is to understand the interplay between scrambling in the stretched horizon and the change in Alice's description of physics as she enters the black hole.
Or, in other words, complementarity maintains that physics as described from inside and outside the black hole should, in an appropriate sense, be equivalent; we want to understand how scrambling---which is a process that occurs from an outside observer's point of view---appears to an observer inside the black hole.

\begin{figure}
\centering
\includegraphics[scale=1]{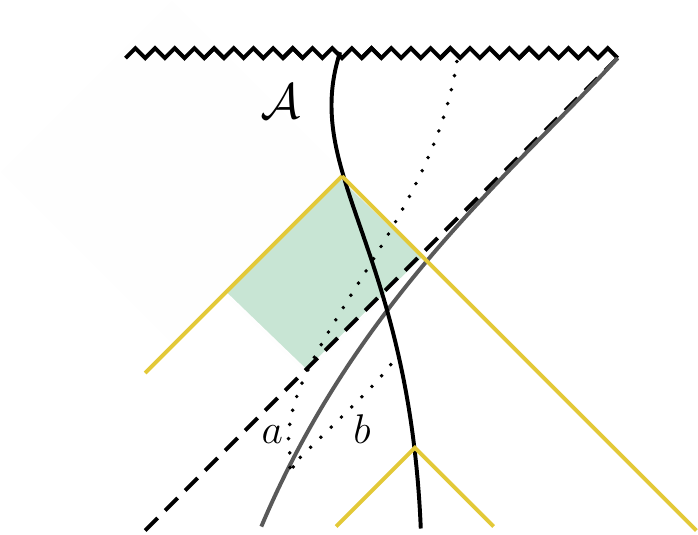}
\caption{Alice's trajectory $\mathcal{A}$ and past lightcones (shown in yellow) as she falls toward the singularity.
The stretched horizon is shown in grey, and the trajectories of the outgoing and ingoing Hawking particles are shown as dotted lines.
We suggest that scrambling causes information about the ingoing excitation to spread out behind the event horizon so that it is delocalized on the intersection of Alice's past lightcones with the causal future of the excitation's horizon crossing point (shaded region).
}
\label{fig:lightcones}
\end{figure}

To be more precise, suppose that Alice follows a timelike trajectory $\mathcal{A}$ that crosses the event horizon and ultimately hits the singularity, as shown in \Fig{fig:lightcones}.
(Partially) foliate the spacetime with her past lightcones.
  When she is inside the black hole, we associate $A$ to the portion of her lightcone that lies inside the black hole.
For all of her lightcones, we associate $O$ to the part of the lightcone that lies outside the black hole and $D$ to the surface where her lightcone intersects the stretched horizon.
According to complementarity, we postulate that for each lightcone whose tip lies inside the black hole, there exists a unitary map
\begin{equation}\label{eq:Ucomp}
\mathcal{U}_\mathrm{comp} \, : \, D \otimes O \; \longrightarrow \; A \otimes O
\end{equation}
that relates the complementary descriptions of physics on either side of the event horizon.
($\mathcal{U}_\mathrm{comp}$ is a effectively a change of basis.)
If scrambling amounts to a unitary process in the stretched horizon, $U_\mathrm{scr} : D \rightarrow D$, then scrambling causes the state of the ingoing modes that Alice finds inside the black hole to evolve according to the action of
\begin{equation} \label{eq:Uin}
\tilde U_\mathrm{scr} \, \equiv \, \mathcal{U}_\mathrm{comp} \left( U_\mathrm{scr} \otimes I_\mathrm{out} \right) \mathcal{U}_\mathrm{comp}^\dagger \, .
\end{equation}

Intuitively, one would expect that scrambling should persist behind the event horizon.
For instance, if one were to drop a qubit into the stretched horizon and wait for it to be well-scrambled, it would be surprising to find it more or less intact and localized after jumping into the black hole.
Moreover, such a discovery would be troubling in light of Hayden and Preskill's finding that the information contained in that qubit is very rapidly returned to the exterior of the black hole \cite{Hayden:2007cs}.
Mathematically, this expectation is equivalent to the statement that we do not expect the unitary operator \eqref{eq:Uin} to act trivially on the physically relevant states in $A$.
We note, however, that it is not logically impossible that $\mathcal{U}_\mathrm{comp}$ exactly undoes the action of $U_\mathrm{scr}$.

On the other hand, it would also be desirable to reconcile the unitary \eqref{eq:Uin} with the semiclassical expectation that spacetime and macroscopic gravitating objects near the event horizon are well-described by general relativity.
Put another way, the field equations of general relativity should be sufficient, at least to a first approximation, to track classical matter thrown into the black hole on timescales where Hawking evaporation is unimportant.
For example, from a semiclassical point of view, if you were to drop a rock into a black hole, you would still expect to find the rock on its freefall trajectory if you accelerated to catch up with it behind the event horizon.

We therefore expect that $\tilde U_\mathrm{scr}$ should act highly nontrivially on fine-grained quantum degrees of freedom, but preserve the coarse-grained state of macroscopically robust and decohered objects. 
More precisely, we expect that the classical geometry inside the black hole should be described by some coarse-graining of $A$, and that the resulting coarse-graining of $\tilde U_\mathrm{scr}$ should act trivially on classical states in this reduced Hilbert space, but that its action on typical states in the full Hilbert space is highly nontrivial.
In particular, this implies that typical ingoing Hawking quanta, which are of course fully quantum, should be rapidly mixed with the rest of the modes in the black hole interior.
On the other hand, a classical observer like Alice should be relatively unaffected by the same dynamics, though of course she will be destroyed in an infall time anyway.
We leave it as an open problem to find a reasonable family of scrambling unitaries that implements little-drama: i.e., dynamics that scrambles small quanta, but leaves classical objects largely intact.
However the arguments that follow will only make use of the fact that the ingoing Hawking quanta are rapidly scrambled over the black hole interior, and not the fact that macroscopic objects are preserved. As such, we will model $U_{\mathrm{scr}}$ (and hence $\tilde U_\mathrm{scr}$) as a generic unitary\footnote{See~\Sec{Sec:Discussion} for a discussion on this simplifying assumption.}.

We emphasize that the dynamics that we have proposed constitute a violation of the no-drama condition, albeit a far milder one than firewalls.
In classical general relativity, the equivalence principle remains intact: the black hole geometry is still described by the Schwarzschild metric, and nothing special happens at the horizon.
Even semiclassically, expectation values of operators should remain unchanged: we are not changing the emission rate of Hawking quanta or the effective temperature of the black hole.
However, working with Hawking emission on a particle-by-particle basis requires a more detailed description. 
We can write the quantum state describing the evaporating black hole in a basis of states which each contain Hawking particles.
In each basis state, individual Hawking quanta are pair-produced as genuine particles (i.e.,\ wavepackets) at a specific spot on the horizon of the black hole, with one wavepacket excitation describing a particle produced in $A$ and a corresponding particle in $O$.
In each basis state, $\tilde U_\mathrm{scr}$ acts to rapidly spread the excitation in $A$ into many other modes, so that after a scrambling time it can no longer be described as a wavepacket or particle.
It is this evolution, which differs dramatically from the propagation of a particle on an empty background metric, that can be seen as violating no drama.

\subsection{Scrambling and kinematics}\label{sub:kinematics}

Next we investigate under what circumstances scrambling of the ingoing modes occurs from Alice's point of view.
Let a clock fixed at the stretched horizon begin ticking when Alice's final particle of Hawking radiation is emitted.
We shall use its reading when Alice reaches the stretched horizon to determine whether or not the corresponding ingoing excitation---which, again, Alice sees as an excitation on the stretched horizon while outside the black hole---has scrambled.

In principle, Alice could wait arbitrarily closely to the stretched horizon so that the ingoing excitation has little time to scramble.
We note, however, that the scrambling time at the stretched horizon is a fantastically small amount of time.
For example, for a supermassive black hole like Sagittarius A* with a mass of about four million solar masses, \Eq{eq:horizonts} predicts that the scrambling time at the stretched horizon should be $\tau_s \approx 3 \times 10^{-42}~\mathrm{s}$, or about 50 Planck times.
As such, Alice does not have much time at all outside of the black hole before scrambling happens, and in practice she will have some amount of computational overhead if she verifies the entanglement between late radiation and early radiation before entering the black hole.
Furthermore, if Alice collects $k>1$ Hawking particles, then scrambling of the first $k-1$ ingoing excitations is virtually guaranteed to have happened before Alice can cross the horizon.
This is because the average rate of Hawking emissions is (much) slower than the rate of scrambling \cite{Page:1976a,Page:1976b}.
Consequently, instances where Alice can cross the horizon before ingoing modes have scrambled are $(k-1)$-fold exponentially suppressed.\footnote{From \cite{Page:1976a}, the cumulative Hawking emission rate for a Schwarzschild black hole is about $10^{-4}~c^3/GM$, so take the characteristic timescale of Hawking emissions to be $t_H \sim 10^4~GM/c^3$.
Note that this is measured in Schwarzschild time, so with the relevant boost factor of $l_P/2r_s$ and for the supermassive black hole discussed above, the characteristic (proper) timescale of Hawking emissions at the stretched horizon is about $(10^3- 10^4)~l_P/c$, which is much larger than the scrambling time.
Also c.f.\ footnote \ref{fn:k} below.}
As we will discuss in the next section, Alice will need to collect $k>1$ Hawking particles in order to be statistically confident in her measurements inside the horizon.

\begin{figure}
\centering
\includegraphics[scale=0.5]{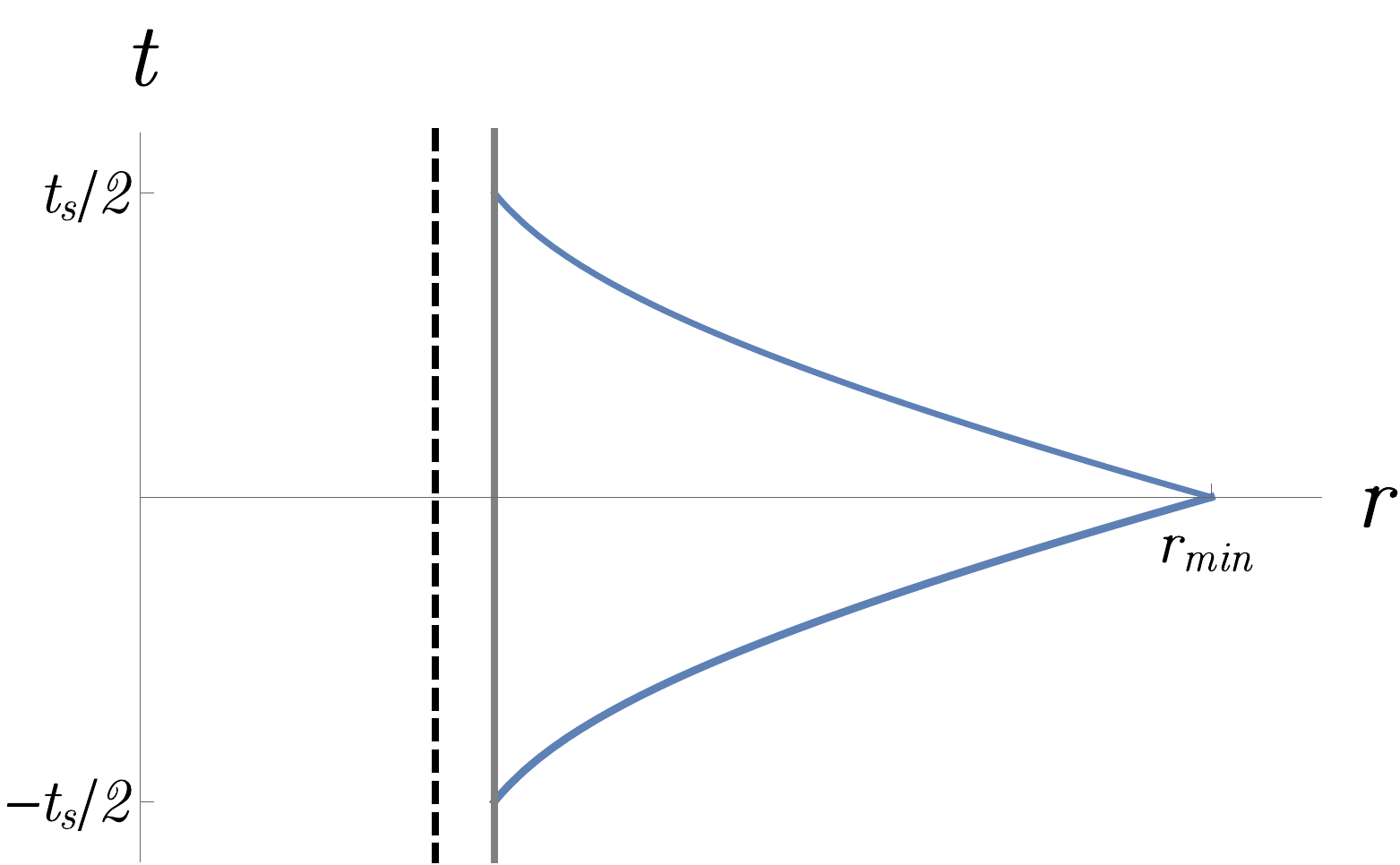}
\caption{Minimum height above which scrambling is guaranteed to occur.}
\label{fig:minheight}
\end{figure}

Separately from the considerations above, it is also interesting to ask what the theoretical minimum height at which Alice can wait above the black hole is above which scrambling is guaranteed to have happened when Alice enters the black hole.
This is the height for which exactly one scrambling time elapses at the stretched horizon in the time it takes a light ray to make a round trip between the stretched horizon and a mirror at the height in question.
This situation is depicted in Schwarzschild coordinates in \Fig{fig:minheight}.

The radial lightlike geodesics are given by
\begin{equation} \label{eq:mingeo}
r - ({r_s}+\delta r) + {r_s} \ln\left[ \frac{{r_s}-r}{{r_s}-({r_s}+\delta r)} \right] = \pm \left(t + \frac{t_s}{2}\right) \, ,
\end{equation}
with $t_s$ and $\delta r$ as given in \Eqs{eq:asympts}{eq:shorcoord} respectively.
The minimum coordinate height is obtained by setting $t=0$ in \Eq{eq:mingeo} and solving for $r$:
\begin{equation}
r_\mathrm{min} = {r_s} \cdot W\left(\frac{\delta r}{{r_s}} \exp\left[\frac{2 \delta r + t_s}{2{r_s}} \right] \right)
\end{equation}
In the above, $W(\cdot)$ denotes the Lambert $W$ function.
The minimum proper distance is therefore given by
\begin{align}
\nonumber \tilde{r}_\mathrm{min} &= \int_{r_s}^{r_\mathrm{min}} \left(1-\frac{{r_s}}{r}\right)^{-1}~dr \\[2mm]
\nonumber &= 2 \sqrt{{r_s} e^{t_s/2R}} \sqrt{\delta r} + O\left((\delta r)^{3/2}\right) \\[2mm]
&\approx \sqrt{\pi} {r_s} \, .
\end{align}
This result is interesting in light of proposals by Nomura, Sanches, and Weinberg \cite{Nomura:2014voa} and by Giddings \cite{Giddings:2015uzr} which both suggest that Hawking radiation is largely invisible to observers unless they are at least on the order of a few Schwarzschild radii away from the horizon of a black hole, which further limits Alice's ability to evade scrambling.

\section{Computation behind the horizon}

To summarize the previous section, if excitations at the stretched horizon are scrambled when Alice reaches the stretched horizon, then we are proposing that the state of the ingoing Hawking modes is thoroughly mixed with other degrees of freedom in the black hole's interior.
In this section we assume that this scrambling has had time to occur; as we explain in \Sec{sub:kinematics}, such a situation should be generic.
As such, Alice is forced to access and process a large number of degrees of freedom that are distributed throughout the interior of the black hole if she wants to verify monogamy of entanglement.
In this section, we discuss how to model the task of verifying entanglement and we investigate its complexity.
In the rest of this paper we will set $l_P = 1$ for brevity. 

\subsection{Model for verifying entanglement}

Following the convention of \cite{Almheiri:2012rt}, we continue to denote the Hilbert space of the interior of the black hole by $A$, and we label the Hilbert spaces of the early radiation and late radiation by $R$ and $B$ respectively (so that $R$ and $B$ are subsets of the space $O$ that we defined in \Sec{sec:scrambling}).
Let $b^{(k)} \subset B$ denote the Hilbert space of the $k$ outgoing Hawking modes that Alice collected and $a^{(k)} \subset A$ the Hilbert space of the corresponding $k$ ingoing modes.
We model $b^{(k)}$ and $a^{(k)}$ each as a collection of $k$ qubits.
Referring to \Eq{eq:Ucomp}, since the Hilbert space $O$ is the same in both complementary descriptions of physics\footnote{We stress, though, that $\mathcal{U}_\mathrm{comp}$ does not factorize over $D$ and $O$.}, it follows that $|A| = |D| = e^{S_{BH}}$, where $S_{BH}$ is the Bekenstein-Hawking entropy of the black hole and where $|\cdot|$ denotes the dimension of a Hilbert space.
As such, we model $A$ as a collection of $n \sim S_{BH}$ qubits that are distributed throughout the interior of the black hole and that are visible to Alice on her past lightcones.

First, what do we mean by ``detecting a violation of the monogamy of entanglement?"
This is nonsensical from the point of view of quantum mechanics, in which monogamy of entanglement is inviolable.
Here, we are given an \emph{apparent} quantum description of entanglement between $b^{(k)}$ and $R$ outside the horizon, and an \emph{apparent} quantum description of entanglement between $b^{(k)}$ and $a^{(k)}$ across the horizon.
While the AMPS paradox shows that there cannot be a global quantum picture that is consistent with both descriptions, the crucial question now is whether Alice can perform an experiment to detect this paradox: in other words, whether she can verify the entanglement between $R$ and $b^{(k)}$, and then verify the entanglement between $b^{(k)}$ and $a^{(k)}$.
If Alice succeeds in verifying both entanglements, then we say that she has detected a violation of monogamy.

What do we mean by verifying entanglement?
In quantum theory, there is no measurement that reliably distinguishes between entangled states and unentangled states---this is because the set of unentangled pure states is non-convex.
However, it is possible to \emph{statistically} test if an unknown state is in a \emph{particular} entangled state. For example, if we let $\ket{\Phi} = \frac{1}{\sqrt{2}} \left( \ket{00} + \ket{11} \right)$ denote an EPR pair, then the two-outcome measurement $M = \{ \ketbra{\Phi}{\Phi}, I - \ketbra{\Phi}{\Phi} \}$ will probabilistically indicate whether a given pair of particles $\ket{\psi}$ is an EPR pair or not.
If $\ket{\psi}$ is indeed an EPR pair, then this measurement will always return outcome $\ketbra{\Phi}{\Phi}$ with certainty.
On the other hand, if $\ket{\psi}$ is an unentangled state $\ket{\phi} \otimes \ket{\theta}$, then it will return outcome $I - \ketbra{\Phi}{\Phi}$ with probability at least $1/2$. While the error of this statistical test is rather large, it can be reduced exponentially by repeating it many times. 
Let $V$ and $W$ denote two disjointed quantum systems.
When we say that Alice has ``verified maximal entanglement between $V$ and $W$,'' we mean that Alice has decoded $k$ pairs of particles from $V$ and $W$, measured each pair using the two outcome measurement $M$, and verified that all $k$ pairs projected to an EPR pair.
This occurs with probability 1 if Alice did indeed decode $k$ EPR pairs; if $V$ and $W$ were unentangled, then this occurs with probability at most $2^{-k}$.
Therefore as $k$ grows, the probability that Alice thinks that $V$ and $W$ are entangled (when they are actually unentangled) becomes exponentially small.
For example, if Alice wants to obtain 5 sigma certainty (error probability $1$ in $3.5$ million) that $V$ and $W$ share maximally entangled particles, she only needs to decode $k=22$ EPR pairs from $V$ and $W$.

\subsection{Alice's computational task}

In this argument, we focus on Alice's task of verifying the entanglement between $b^{(k)}$ and $a^{(k)}$ when she jumps into the black hole---we will assume that she has already verified the entanglement between $b^{(k)}$ and $R$ prior to jumping in. We consider the quantum description of the black hole interior $A$, along with the late-time Hawking modes $b^{(k)} a^{(k)}$. 
Consider the moment at the stretched horizon that $k$ Hawking pairs $b^{(k)} a^{(k)}$ were produced\footnote{For simplicity here we assume that they are produced simultaneously rather than one-by-one, but this does not hinder the argument. Indeed, if they are produced sequentially, then due to arguments by Page \cite{Page:1976a,Page:1976b}, the average rate of Hawking pair production is less than one pair per scrambling time.
Therefore, in a sequential production picture, all but the last Hawking pair will have been scrambled by the time that Alice can enter the black hole.
If Hawking radiation can be modeled thermally, sequential emission is exponentially preferred over simultaneous emission.\label{fn:k}}.
The state of the Hawking pairs and the black hole interior can be described by the density matrix
$$
	\sigma^{b^{(k)} a^{(k)} A} = (\ketbra{\Phi}{\Phi}^{\otimes k})^{b^{(k)} a^{(k)}} \otimes \rho^A \, ,
$$
where $\ket{\Phi} = \frac{1}{\sqrt{2}} \left( \ket{00} + \ket{11} \right)$ is a maximally entangled Hawking pair, and $\rho^A$ is the density matrix of the black hole interior right before the pair production event.
By Page's theorem \cite{Page:1993,Page:2013}, after the Page time $\rho^A$ is close to being maximally mixed; for the remainder of this argument, we will assume that $\rho^A$ is exactly the maximally mixed state on $n$ qubits.\footnote{If $\rho^A$ is $\eps$-close to the maximally mixed state, then our final bounds will only acquire an additional $\eps$ additive error.} 

As discussed in the previous section, by the time that Alice arrives at the stretched horizon with $b^{(k)}$ in tow, the black hole interior (which now includes $a^{(k)}$) has experienced extensive scrambling. We model this as follows.
Let $U$ be the unitary representing the scrambling dynamics, which acts on $A' = a^{(k)} A$.
From Alice's point of view, the state of the scrambled interior $A'$ and $b^{(k)}$ can then be described by
$$
	\tau^{b^{(k)} A'} = (I^{b^{(k)}} \otimes U^{A'}) \sigma^{b^{(k)} A'} (I^{b^{(k)}} \otimes U^{A'})^\dagger .
$$
Because our understanding of the quantum mechanical evolution of black holes is rather limited, we will model the unitary $U$ as being Haar-random.
(In fact our arguments will carry through in the case that $U$ is chosen from an ensemble of efficiently constructible unitaries that is sufficiently randomizing; we will discuss this in more detail in~\Sec{Sec:Discussion}.)

As Alice falls towards the singularity, she attempts to interact with a set $S$ of qubits of the interior in order to recover at least one unit of entanglement between the interior and $b^{(k)}$.
First, suppose $S$ is a subsystem of $A'$ that has at most $n/2$ qubits.
Then, by~\cite{Hayden:2007cs}, we have that
\begin{equation} \label{eq:decoupling}
	\int dU \left \| \tau^{b^{(k)} S} - \tau^{b^{(k)}} \otimes \tau^{S} \right \|_1^2 \leq |b^{(k)} S| \cdot \Tr \left [ (\sigma^{b^{(k)} A'})^2 \right ].
\end{equation}
We have that $\Tr \left [ (\sigma^{b^{(k)} A'})^2 \right ] = \Tr \left [ (\ketbra{\Phi}{\Phi}^{b^{(k)} a^{(k)}})^{\otimes k} \otimes (\rho^A)^2 \right ] = \Tr[(\rho^A)^2] = 2^{-n}.$ The dimension of $b^{(k)} S$ is at most $2^{n/2 + k}$, so therefore 
$$
\int dU \left \| \tau^{b^{(k)} S} - \tau^{b^{(k)}} \otimes \tau^{S} \right \|_1^2 \leq 2^{-n/2 + k}.
$$
Thus, by the time Alice reaches the event horizon, with probability exponentially close to one (over the choice of unitary $U$), any subset $S$ of at most $n/2$ qubits of the interior of the black hole will essentially be uncorrelated with her Hawking modes $b^{(k)}$: the black hole dynamics ``smears'' the entanglement between $b^{(k)}$ and $a^{(k)}$ over the entirety of the black hole. This holds for as long as $k \ll n/2$, i.e., as long as the amount of material that Alice brings with her into the black hole is negligible compared to the size of the black hole\footnote{Otherwise, if Alice is bringing a sizable fraction of the black hole's mass with her across the horizon, this could plausibly take the state of the black hole to before the Page time, change the horizon size, or any number of other nonperturbative effects which break the setup of the paradox.}.
Therefore, unless Alice interacts with more than half of the qubits of the black hole, she has no hope of decoding a partner qubit that is maximally entangled with $b^{(k)}$ after crossing the event horizon. 

However, can Alice interact with more than half of the qubits in $A^\prime$?
We assume that Alice is a localized experimenter (such that she is unable to do parallel computation on a spacelike region), so that she can only process at most $O(1)$ qubits of the black hole interior per Planck time.
Thus, to touch at least $n/2$ qubits, Alice would require $\Omega(n)$ Planck times.
However, Alice also has no chance of doing this before experiencing an untimely demise: the longest amount of time that can elapse on Alice's clock before she reaches the singularity is $O({r_s})=O(\sqrt{n})$ in Planck units.
Again, she has no hope of decoding any entanglement between $b^{(k)}$ and $A'$.
In other words, because of black hole scrambling, Alice does not have enough time to verify the entanglement between $b^{(k)}$ and $a^{(k)}$, and thus is unable to perform the AMPS experiment.

\section{Discussion}
\label{Sec:Discussion}

We now elaborate upon several aspects of our argument, including discussing possible objections.

\subsection{Modeling scrambling dynamics}
\label{Sec:scrambling_model}
In our argument, we model the scrambling dynamics of the black hole as a generic unitary sampled from the Haar distribution. As mentioned before, we model $U_{scr}$ as a generic unitary in order to capture the part of little-drama where fine-grained quanta get scrambled. It does not model the other part of little-drama where macroscopic objects are preserved, but we do not use this part in our argument. 

An immediate objection to this modeling choice is that black hole dynamics cannot, strictly speaking, look anything like a Haar-random unitary. This is because a generic unitary will have exponential complexity: the minimum number of local quantum operations that need to be applied in order to implement the unitary -- known as the \emph{circuit complexity} of the unitary -- is exponential in the number of its degrees of freedom. Assuming the Physical Church-Turing Thesis\footnote{Briefly, the Physical Church-Turing Thesis states that all computations in the physical universe can be simulated, with polynomial time overhead, by a universal quantum computer.}, an $n$-qubit black hole that evolves for $\mathrm{poly}(n)$ Planck times should only be able to realize unitaries that have $\mathrm{poly}(n)$ circuit complexity, where $\mathrm{poly}(n)$ denotes some polynomial in $n$. Perhaps unitary matrices with polynomial circuit complexity will not adequately ``smear'' entanglement across the entire black hole interior, as required by our argument. 

As noted by Hayden and Preskill~\cite{Hayden:2007cs}, one can model the dynamics of a black hole using \emph{random unitary designs}. Informally speaking, unitary designs are ensembles of unitaries with polynomial circuit complexity that in many respects behave like Haar-random unitaries. In our argument, the Haar unitary ensemble can be replaced by an (approximate) unitary design and our conclusion remains essentially unchanged: unitary designs, though possessing small circuit complexity, still ``smear'' quantum information across all degrees of freedom. Unitary designs have been extensively studied in the quantum information community. By now, we know several examples of (approximate) unitary designs~\cite{brandao2012local,dankert2009exact}.

Still, what do we mean when we say that a \emph{particular} black hole behaves like a unitary randomly chosen from an ensemble? After all, a black hole behaves according to none other but \emph{the} unitary given by the theory of quantum gravity. Unfortunately, since this theory is still unavailable to us, in our calculations we must make a ``best guess'' at what a black hole unitary must look like. Without presupposing unjustified constraints on the theory of quantum gravity, our best guess for black hole dynamics is that the Hamiltonian governing the interior should be local and strongly mixing, and that the black hole evolves in polynomial time. The Maximum Entropy Principle from statistics and learning theory tells us that our best guess for the black hole unitary is a randomly chosen one from the uniform distribution over unitaries with polynomial circuit complexity\footnote{The Maximum Entropy Principle is a formalization of Occam's Razor in machine learning and statistical learning theory~\cite{shore1980axiomatic}. It says that, given a set of hypotheses consistent with one's observations, one's best hypothesis is the maximum entropy one: a randomly chosen one from that set.}. We note that this ensemble of unitaries is known to form an approximate unitary design~\cite{brandao2012local}, and thus has the scrambling properties required by our argument.

\subsection{Black holes in other dimensions.} 
\label{subsec:complexity} One may also object that this argument is specific to spacetimes of dimension 3+1.
In higher dimensions this argument only becomes stronger, since in spacetimes with spatial dimension $d$, the number of qubits that make up the interior Hilbert space, $|A|$, scales like $O(r_s^{d-2})$, while the infall time scales like $O({r_s})$.
As such, the infall time is increasingly smaller with respect to $|A|$ for $d > 3$.
But, this is not necessarily true for lower spatial dimensions.
For example, in $\mathrm{AdS}_3$, the number of qubits and the infall time both scale linearly with ${r_s}$.
Consequently, our previous trivial bound on the number of accessible qubits does not suffice here.
In this case one can appeal to the fast scrambling conjecture to render the computation impossible.
The fast scrambling conjecture of Sekino and Susskind \cite{Sekino:2008he} states (among other things) that black holes are the fastest scramblers in nature\footnote{We note that the fast scrambling conjecture stating that the fastest scrambling time for a black hole is ${r_s} \log {r_s}$ is an asymptotic statement, and thus not broken by earlier statements of $\log {r_s}$ scrambling time at the stretched horizon.}.
Lashkari \emph{et al.} \cite{Lashkari:2011yi} formalized this notion in terms of quantum information by stating that black holes saturate the ${r_s}\log {r_s}$ lower bound for scrambling time.
In this work, we consider a quantum complexity formulation:
\begin{conjecture} Let $k \ll n/2$, i.e., let $k$ be much smaller than the number of qubits in the black hole. Let $U$ be the unitary corresponding to running black hole dynamics for time $t$ on $A'=a^{(k)}A$, as measured by an asymptotic observer.
Then recovering the entanglement between $a^{(k)}$ and $b^{(k)}$ from $A'$ and $b^{(k)}$ requires time at least $t$.
More formally, for any unitary $V$ acting on system $A'$, if $\nu^{b^{(k)} A'} = (I^{b^{(k)}} \otimes V^{A'}U^{A'}) \sigma^{b^{(k)} A'} (I^{b^{(k)}} \otimes V^{A'} U^{A'})^\dagger$ is the state of the system after applying $VU$ to $A'$ , and if 
\[\left \| \nu^{b^{(k)} A'} - \nu^{b^{(k)}} \otimes \nu^{A'} \right \|_1^2 \geq \delta \, , \]
where $\delta$ is a small constant (say 0.01), then $V$ has circuit depth at least $t$.
\end{conjecture}

This is a circuit-depth version of the statement ``black holes are the fastest scramblers in nature."
It says that if one wishes to invert the scrambling performed by the black hole, then one requires at least the scrambling time to do so.
If such a statement is true, then in our model, unscrambling the entanglement between $a^{(k)}$ and $b^{(k)}$ requires at least ${r_s}\log {r_s}$ time in any dimension, whereas the infall time scales as ${r_s}$.
Therefore, such a conjecture would suffice for our arguments to hold in any dimension.

\subsection{Localization of the experimenter.} 

In our argument, we assume that Alice is localized throughout our experiment, and therefore can access only $O(r_s)$ qubits after crossing the horizon. One might object that if one knew the exact dynamics of $\tilde{U}_\mathrm{scr}$, one could set up the infalling matter such that a nonlocal experiment is performed on the interior modes and the result is then sent to Alice. However, this is impossible because Alice is out of causal contact with most of the black hole interior \cite{Freivogel} from which the results of the nonlocal experiment would have to be sent. 
Therefore, even this non-local experiment cannot reveal entanglement between the interior and exterior Hawking modes before Alice hits the singularity.

\subsection{Relation to prior works}
\label{sec:priorworks}

We first note that in \cite{Freivogel} arguments have already been made about the inability of the infalling observer to access the entirety of the interior of the black hole except at the singularity. These arguments are quite different in nature from the information-theoretic ones of this paper. In particular, there appears to be the possibility to work around the arguments in \cite{Freivogel} by using multiple observers \cite{RajuCausalPatches}, something which does not seem to be an issue in the more information-theoretic arguments of this note.

Readers may notice that our argument significantly resembles that given by Hayden and Preskill \cite{Hayden:2007cs}.
While the techniques are similar, our conclusions and assumptions differ in several ways.
First, \cite{Hayden:2007cs} concludes that black holes, rather than being information sinks, are plausibly more like information ``mirrors;" information deposited into the black hole gets released (in scrambled form) as quickly as possible.
On the other hand, our goal is to demonstrate a \emph{lower bound} on Alice's ability to recover a single qubit of information within the black hole after it has been scrambled.
Second, Hayden and Preskill explicitly model the joint state of the black hole, its radiation, as well as some reference system as a pure state.
However, in the context of the firewalls paradox, we cannot write down such a description to begin with!
In our setting, we focus solely on the part of the black hole that Alice sees after she has collected her Hawking mode and has crossed the event horizon.
This is consistent with complementarity; we only need to provide a valid description of physics inside the horizon, which need not be in a tensor product with the description of physics outside the horizon.

Our proposal also shares some spiritual similarities with fuzzball complementarity \cite{Mathur:2013gua}, in which undisturbed freefall through the horizon is recovered in the limit where the incident energy of the observer is much larger than the temperature of the black hole, in the sense that local properties of the infalling observer are important to consider in both cases. 
We note that in the context of the fuzzball program, the definition of complementarity invoked by AMPS---which we follow in \Sec{sec:scrambling} when we define the Hilbert space relevant to the problem---is replaced by a different and perhaps more correct definition involving the definition of the state along the complete slice, both inside and outside of the horizon. 
While it would be interesting to reformulate our results in that lens, it is perhaps unnecessary: in that limit the fuzzballs program already precludes the need for a different resolution to the information paradox! 
Instead, we emphasize that, even when cleaving as close to AMPS-style complementarity definitions as possible, information- and complexity-theoretic arguments by themselves strongly constrain the ability for any observer to actually observe violation of monogamy of entanglement.

We also differ from the fuzzballs approach in analyzing operationally what is possible for the observer to compute after crossing the stretched horizon of the black hole on the way to an existent singularity.
In this work, the singularity plays a vital role in determining the longest possible time available to perform the computation.
But, in fuzzball complementarity, the singularity is fuzzed out and resolved at some characteristic fuzzball radius, behind which space stops existing. 
It may be interesting to see by what degree our bounds would tighten in the specific case of fuzzballs; we reiterate, though, that we are already able to demonstrate that we cannot operationally detect monogamy of entanglement even without the shorter longest possible time for the computation given by the fuzzball program.

Finally, we also note the recent paper \cite{Mathur:2015nra}, which provides a concrete toy model for fuzzball complementarity.
It would be interesting to examine our proposals in the context of this work, since the dynamics of infalling excitations discussed in \cite{Mathur:2015nra} may be able to inspire and inform a similarly concrete realization of the scrambling dynamics that we discussed in \Sec{sec:innout}.

\subsection{Other black hole geometries}

We have thus far restricted our attention to only Schwarzschild black holes. It is a reasonable question to ask what happens once we consider other geometries with nonzero spin or charge. With regard to these, the addition of spin or charge to a black hole splits the horizon into an inner and an outer horizon.
It is possible in such geometries to spend a longer amount of time between the two horizons, so in principle Alice could have enough time to complete her monogamy verification before hitting the singularity, thus implying a naive breakdown of the story up to this point.
Alternatively, in maximal extensions of these black hole spacetimes, Alice could pass from the black hole interior into other asymptotically flat spacetime regions and continue to exist indefinitely.

We note, however, that the inner horizon is not entirely understood, both from the perspective of general relativity and quantum theory \cite{Dotti:2010uc, Engelhardt:2015gla}.
(For example, the inner horizon is strongly believed to be unstable.)
As such, it is likely that our assumptions about quantum mechanics and general relativity would need to be modified (at least in the vicinity of the inner horizon) in order to discuss charged spinning black holes, and it is another question entirely what form the AMPS paradox would take if it persists.

\section{Conclusion}

We have described a resolution of the information paradox that amounts to a weakening of the no-drama condition --- a new condition that we call little-drama.
We suppose that quantum systems that cross the event horizon of a black hole experience nontrivial evolution which entangles them with other degrees of freedom in the black hole interior.
Such evolution inside the horizon is the complementary description of scrambling on the stretched horizon and constitutes a mild departure from the predictions of a non-gravitating field theory.

The little-drama condition allows for an apparent violation of monogamy of entanglement that is similar in spirit to the Harlow-Hayden proposal.
Past the Page time, an observer can verify that early and late Hawking radiation have the right entanglement structure outside of a black hole and then smoothly pass through the event horizon.
While the smooth crossing implies a violation of monogamy of entanglement---it would seem that the late radiation is maximally entangled with both the early radiation and the black hole interior---we found that the observer could not verify this violation before encountering the singularity.

It is also worth emphasizing that, as an information-theoretic proof, our arguments for larger than three spacetime dimensions are resilient to the Oppenheim-Unruh precomputation-style attacks, which are complexity-theoretic in nature. 
Though our complexity-theoretic argument (which holds in all dimensions) does not necessarily share this feature, it is possible that precomputation cannot simultaneously prevent both our construction and the Harlow-Hayden argument from resolving the AMPS paradox.
Two distinct and mutually exclusive precomputation style attacks are required to foil both obstacles to AMPS.
In the first, one collapses halves of Bell pairs into a black hole to evade Harlow-Hayden.
In the second, one takes entire Bell pairs and collapses them into a black hole to evade our arguments.
We note it is not simultaneously possible to do both for any single qubit.
Therefore these two resolutions of the information paradox might be complementary in a different sense of the word.

Directions for future research include finding a model for black hole dynamics that faithfully captures all parts of little-drama. Other directions include working out the details for other black hole geometries with nonzero spin or charge.
As previously discussed, it is not clear that such geometries would be precluded from violation of monogamy of entanglement in the same way, but a parametric comparison of how much leeway they have would be interesting to conduct.
It would also be interesting if the information-theoretic proof method could be extended to spacetimes with fewer than three spatial dimensions without assuming the fast-scrambling conjecture.

\vspace{10pt}
\medskip
\medskip

\noindent \textbf{Acknowledgements.} We thank Wilson Brenna, Charles Cao, Sean Carroll, Daniel Harlow, Jonathan David Maltz, Grant Remmen, and Douglas Stanford for helpful discussions.
This is material is based upon work supported in part by the following funding sources:
N.B. is supported in part by the DuBridge Postdoctoral Fellowship, by the Institute for Quantum Information and Matter, an NSF Physics Frontiers Center (NFS Grant PHY-1125565) with support of the Gordon and Betty Moore Foundation (GBMF-12500028), and by the U.S. Department of Energy, Office of Science, Office of High Energy Physics, under Award Number DE-SC0011632.
A.B. is supported in part by the NSF Graduate Research Fellowship under grant no. 1122374 and by the NSF Alan T. Waterman award under grant no. 1249349.
A.C.-D. is supported by the NSERC Postgraduate Scholarship program.
J.P is supported in part by DOE grant DE-SC0011632 and by the Gordon and Betty Moore Foundation through Grant 776 to the Caltech Moore Center for Theoretical Cosmology and Physics.
H.Y. was supported by Simons Foundation
grant 360893, and National Science Foundation grant 1218547.

\end{document}